\documentstyle[12pt]{article}

\begin{document}

\hfill{UTTG-08-03}

\vspace{36pt}

\begin{center}
{\large {\bf { Can Non-Adiabatic Perturbations Arise After Single-Field Inflation? }}}

\vspace{36pt}
Steven Weinberg\footnote{Electronic address:
weinberg@physics.utexas.edu}\\
{\em Theory Group, Department of Physics, University of
Texas\\
Austin, TX, 78712}

\vspace{30pt}

\noindent
{\bf Abstract}
\end{center}

\noindent
It is shown that non-adiabatic cosmological perturbations cannot appear during the period of reheating following inflation with a single scalar inflaton field.  

 \vfill

\pagebreak
\setcounter{page}{1}

According to a widely adopted picture [1], the perturbations to the Robertson--Walker cosmology arose from the quantum fluctuations in a slowly rolling scalar ``inflaton'' field during a period of inflation, then became classical as their wavelength was stretched beyond the horizon, and subsequently were imprinted on the decay products of the inflaton during a period of ``reheating.''  One of the attractions of this picture (and in particular the assumption that there is just one inflaton field) is that it has generally been thought to lead only to adiabatic perturbations, in agreement with current observations [2].  

A recent preprint [3] has raised the question, whether it is possible for non-adiabatic cosmological perturbations to arise during reheating even after inflation with a single inflaton field.  This would be very important if true, for then observational limits on non-adiabatic fluctuations in the cosmic microwave background would  
provide some constraints on the otherwise mysterious era of reheating, and indeed on the whole
history of the universe between inflation and the present.  

However, there are very general grounds for expecting that single-field inflation can only produce adiabatic fluctuations, whatever happens in reheating or subsequently. It has been shown [4] that, whatever the constituents of the universe, the differential equations for cosmological perturbations in Newtonian gauge {\em always} have an adiabatic solution for wavelengths outside the horizon (that is, for physical wave numbers that are much less than the cosmological expansion rate).  For this solution, there is a conserved quantity $\zeta$ [5], defined by
\begin{equation} 
\zeta\equiv-\Psi+\delta\rho/3(\bar{\rho}+\bar{p})\;,
\end{equation} 
where $\delta\rho$ is the perturbation to the total energy density in Newtonian gauge; 
bars denote unperturbed quantities; 
the perturbed metric is given by
\begin{equation} 
ds^2=-(1+2\Phi)dt^2+a^2(1-2\Psi)\,d{\bf x}^2\;;
\end{equation} 
and as usual $H\equiv \dot{a}/a$.  (Reference [4] dealt mostly with a quantity [6]
\begin{equation}
{\cal R}\equiv -\Psi+H\,\delta u\;.
\end{equation} 
where $\delta u$ is the perturbation to the total velocity potential, but outside the horizon ${\cal R}$ and $\zeta$ are the same.)   Also, for this adiabatic mode,
the perturbations to the metric and the total energy density and pressure are given outside the horizon by 
\begin{eqnarray}
&&\Phi=\Psi=\zeta\left[-1+\frac{H(t)}{a(t)}\int_{t_1}^t a(t')\,dt'\right] \\
&&\delta\rho=-\frac{\zeta\dot{\bar{\rho}}(t)}{a(t)}\int_{t_1}^t a(t')\,dt'\;,~~~~~~
\delta p=-\frac{\zeta\dot{\bar{p}}(t)}{a(t)}\int_{t_1}^t a(t')\,dt'\;.
\end{eqnarray} 
This is a solution for any value of $t_1$, so the difference of adiabatic solutions with different values of $t_1$
is also a solution, with $\zeta=0$ and $\Phi=\Psi\propto H(t)/a(t)$, $\delta\rho/\dot{\bar{\rho}}\propto 1/a(t)$, etc.   We will adjust $t_1$ so that the total perturbation takes the form (4), (5), in which case $t_1$ will be at some early time during inflation to avoid perturbations that are very
large at early times. 
Furthermore, if the total energy-momentum tensor $T^{\mu\nu}$ is given by a sum of tensors 
$T_\alpha^{\mu\nu}$ for fluids labeled $\alpha$ (not necessarily individually conserved) then Eq.~(5) holds for each of the
individual perturbations 
\begin{equation} 
\delta\rho_\alpha=-\frac{\zeta\dot{\bar{\rho}}_\alpha (t)}{a(t)}\int_{t_1}^t a(t')\,dt'\;,~~~~~~
\delta p_\alpha =-\frac{\zeta\dot{\bar{p}_\alpha }(t)}{a(t)}\int_{t_1}^t a(t')\,dt'\;,
\end{equation}  
and similarly for any other four-dimensional scalar, such as the inflaton field
\begin{equation} 
\delta\varphi=-\frac{\zeta\dot{\bar{\varphi}}(t)}{a(t)}\int_{t_1}^t a(t')\,dt' \;.
\end{equation} 

In general there may be other solutions, for which $\zeta$ may or may not be constant and the perturbations are not given by Eqs.~(4)--(7).    (Misleadingly, these others are often called isocurvature solutions.)  But if only one inflaton field makes a contribution to the energy-momentum tensor during inflation then the perturbations to this field and the metric in Newtonian gauge are governed by a third-order set of differential equations with a single constraint, so only two independent solutions contribute to cosmological perturbations, and since we have found two explicit adiabatic solutions outside the horizon, these are the only solutions for the coupled system of inflaton and gravitational fields from horizon exit to the end of inflation.  During reheating and the subsequent evolution of the universe other fields and fluids become important, but the adiabatic mode is a true solution outside the horizon whatever the contents of the universe, so with the universe in the adiabatic mode at the beginning of reheating it remains in this mode as long as the wavelength remains outside the horizon.

So how did non-adiabatic modes arise in reference [3]?  The theorem of reference [4] assumes that the field equations hold at every moment, which requires that the perturbations are differentiable functions of time.  As he suggested might be the case, the reason that Armend\'{a}riz-Pic\'{o}n found a non-adiabatic mode 
during reheating in reference [3] was that, although it was assumed that initially during inflation only the inflaton field was present and there was no energy transfer to other fields, he supposed in this reference that the energy transfer rate rose discontinuously to a non-zero value at the beginning of reheating.  (The model considered in reference [3] actually gave a constant value for $\zeta$,
but the perturbations had unequal values of  $\delta\rho/\dot{\bar{\rho}}$ for the inflaton and its decay products, so as recognized by Armend\'{a}riz-Pic\'{o}n, this perturbation was not adiabatic.)
Of course, a discontinuous change in the energy transfer rate is unphysical.

There is  one weak point in the above general argument that non-adiabatic perturbations do not arise during
reheating.  It  is the assumption that there was nothing but the inflaton and gravitation before reheating.  Of course, in order for reheating to occur at all, there must be other fields or fluids besides the inflaton, and these do not suddenly come into existence during reheating.  The reactions that
produce matter during reheating can't be completely absent beforehand, so the remaining question  is whether the transfer of energy from the inflaton to other fields during inflation before reheating excites these other modes.

This question is answered by a further theorem, that if the matter energy density during inflation is small, then even if the perturbations to the matter energy density are initially not at all adiabatic, the departures from adiabaticity would decay exponentially fast as soon as the energy transfer rate becomes appreciable. 

Here is the proof.  The  co-moving rate per proper volume of energy transfer from the inflaton field $\varphi$ to  ``matter'' fields (possibly including radiation) is in general some four-scalar function $X$ of all these fields and perhaps their first and higher time derivatives:
\begin{equation}
-u_\mu {T^{\mu\nu}_M}_{;\nu}=X(\varphi,\dots)\;,
\end{equation} 
where $u^\mu$ is the four-vector velocity of the total energy-momentum tensor, normalized so that $u^\mu u_\mu=-1$.  To zeroth order in all perturbations to the metric and other fields, this reads
\begin{equation} 
\dot{\bar{\rho}}_M+3H(\bar{\rho}_M+\bar{p}_M)=\bar{X}\;,
\end{equation} 
where bars denote unperturbed values, taken to depend only on time.  (We choose signs so that $\bar{u}^0=+1$.)  To first order in perturbations,
Eq.~(8) gives in Newtonian gauge
\begin{equation} 
\delta\dot{\rho}_M+3H(\delta\rho_M+\delta p_M) -3(\bar{\rho}_M+\bar{p}_M)\dot{\Psi}=\delta X+\Phi \bar{X}\;,
\end{equation} 
the last term on the right arising from the perturbation to $u_0$.
We have dropped terms involving spatial gradients, which become negligible outside the
horizon.  

Now suppose that  at some early time during inflation the density and pressure of matter are  small.  (This is plausible, because the energy density of fermions and gauge fields produced by quantum fluctuations would be quadratic in the fluctuations.)  
Then the inflaton will provide the chief source of the gravitational field.  As mentioned above, under these conditions the perturbations to the inflaton and gravitational potentials will be described by the adiabatic solution
\begin{equation} 
\delta\varphi=-\dot{\bar{\varphi}}{\cal I}\;,~~~~~~~\Phi=\Psi=-\zeta+H{\cal I}\;,
\end{equation} 
in which for convenience we have  introduced the notation
\begin{equation} 
{\cal I}(t)\equiv \frac{\zeta}{a(t)}\int_{t_1}^t a(t')\,dt'\;.
\end{equation} 
Also, under these conditions the energy transfer rate will depend only on the inflaton and gravitational fields and their time derivatives, so
\begin{equation} 
\delta X = -\dot{\bar{X}}{\cal I} \;.
\end{equation} 
(For instance, if $X$ depends only on $\varphi$, then 
$$\delta X=\frac{\partial \bar{X}}{\partial \bar{\varphi}}\delta\varphi=-\left(\frac{\partial \bar{X}}{\partial \bar{\varphi}}\dot{\bar{\varphi}}
\right)\,{\cal I} \;.
$$
Eq.~(13) also holds if $X$ depends also on $\dot{\varphi}$, $\ddot{\varphi}$, etc., provided
that for each pair of time derivatives there is a factor of $g^{00}$ to keep $X$ a scalar.)
Putting Eq.~(13) together with Eq.~(11), the right-hand side of Eq.~(10) is
\begin{equation}
\delta X+\Phi \bar{X}=- \frac{\partial}{\partial t}\Big[\bar{X}{\cal I}\Big]\;.
\end{equation} 
and the difference between Eq.~(10) and the time-derivative of Eq.~(9) gives
\begin{equation} 
\frac{\partial}{\partial t}\left[\delta\rho_M+\dot{\bar{\rho}}_M{\cal I}\right]
=-3H(\delta\rho_M+\delta p_M)-3(\bar{\rho}_M+\bar{p}_M)\ddot{{\cal I}}
-3\frac{\partial}{\partial t}\left[\Big(\bar{\rho}_M+\bar{p}_M\Big)H{\cal I}\right]\;.
\end{equation} 
This can be simplified by noting that ${\cal I}$ satisfies the differential equation
\begin{equation}
\ddot{{\cal I}}+\frac{\partial}{\partial t}(H{\cal I})=0\;,
\end{equation} 
so
\begin{equation} 
\frac{\partial}{\partial t}\left[\delta\rho_M+\dot{\bar{\rho}}_M{\cal I}\right]
=-3H(\delta\rho_M+\delta p_M+\dot{\bar{\rho}}_M{\cal I}+\dot{\bar{p}}_M{\cal I})\;.
\end{equation}

To continue, let us assume  that the matter pressure  is a function
$p_M(\rho_M)$ only of the matter energy density $\rho_M$, as in the case of pure radiation
or pure dust.  (We are {\em not} assuming this
for the combined system of matter and inflaton.)  This is plausible, because the decay of a single
real inflaton field would not generally produce any chemical potentials.  Then $\dot{\bar{p}}_M=p_M'(\bar{\rho}_M) \dot{\bar{\rho}}_M $ and $\delta p_M=p_M'(\bar{\rho}_M)\delta \rho_M$.  Eq.~(17) now reads
\begin{equation} 
\frac{\partial}{\partial t}\left[\delta\rho_M+\dot{\bar{\rho}}_M{\cal I}\right]
=-3H(1+ c_M^2 )(\delta\rho_M+\dot{\bar{\rho}}_M{\cal I})\;,
\end{equation}
where $c_M^2\equiv d p_M/d\rho_M $ is the squared sound speed.
This shows that $|\delta\rho_M+\dot{\bar{\rho}}_M{\cal I}|$ decreases monotonically
and faster than $a^{-3}$, but that is not good enough, because we have to make sure
that this  is not just because $|\delta\rho_M|$ and $|\dot{\bar{\rho}}_M{\cal I}|$
are both decreasing.  For this purpose, we use Eq.~(9) again, and re-write Eq.~(18) as
\begin{equation}
\frac{\partial}{\partial t} \ln N= -\bar{X} \frac{(1+ c_M^2 )}{\bar{\rho}_M+\bar{p}_M}\;,
\end{equation} 
where $N$ is a dimensionless measure of the departure of the matter energy density perturbation from its adiabatic value
\begin{equation}
N\equiv \left|\frac{\delta\rho_M+\dot{\bar{\rho}}_M{\cal I}}{\bar{\rho}_M+\bar{p}_M}\right|\;.
\end{equation} 
Since energy is flowing from the inflaton field to matter, $X$ is positive, and Eq.~(19)  shows that $N$ decreases monotonically.  Early in inflation the matter  perturbation may be nowhere near adiabatic, with $N$ of the same order of magnitude as the fractional density perturbation.  As time passes during inflation, the transfer of energy from the inflaton to matter may make both $\delta\rho_M$ and $\dot{\bar{\rho}}_M$ large, but the quantity $N$ continues to decrease.  Eventually, after the energy transfer rate $X$ becomes large for a sufficiently long time during reheating, the matter density perturbation can no longer 
be ignored in calculating  $X$ and the gravitational field perturbations,
but by that time $N$ will have decayed exponentially.   With the  density and pressure perturbations of the matter as well as the inflaton satisfying Eq.~(6), we will still have
$\delta X=-\dot{\bar{X}}{\cal I}$ and $\Phi=\Psi=-\dot{{\cal I}}$, and the above analysis will
remain valid.  

Note that asymptotically, after the transfer of energy from the inflaton to matter
ceases, the quantity $-\dot{\bar{\rho}}_M{\cal I}/(\bar{\rho}_M+\bar{p}_M)$ approaches $3\zeta/(1-\dot{H}/H^2)$, so as long as $\bar{p}_M<\bar{\rho}_M$, $-\dot{\bar{\rho}}_M{\cal I}/(\bar{\rho}_M+\bar{p}_M)$ is bounded below in absolute value by
$3|\zeta|/4$.  Thus the reason that $N$ approaches zero is {\em not} that $\delta\rho_M/(\bar{\rho}_M+\bar{p}_M)$
and $-\dot{\bar{\rho}}_M{\cal I}/(\bar{\rho}_M+\bar{p}_M)$ are both approaching zero.  Rather, the ratio of these quantities, $-\delta\rho_M/\dot{\bar{\rho}}_M {\cal I}$, approaches unity.

With the matter pressure a function only of the
matter density, the matter pressure perturbation will then also satisfy Eq.~(6).  Thus the 
perturbations become adiabatic, in the sense of Eq.~(6) and (7).  Also, 
with the energy density and pressure perturbations of  matter
as well as of the inflaton all satisfying Eq.~(6),  the total energy density and 
pressure perturbations will obviously satisfy Eq.~(5).  It follows that the perturbations 
also satisfy the adiabatic condition  that $\zeta$ is constant,
because in general 
\begin{equation} 
\dot{\zeta}=\frac{\dot{\bar{\rho}}\,\delta p-\dot{\bar{p}}\delta\rho}{3(\bar{\rho}+\bar{p})^2}\;.
\end{equation}  
The same analysis also obviously applies if the inflaton energy goes into several species 
of matter, with a pressure for each species given by a function  of the energy density in that species. 

In conclusion, even if the decay of the inflaton during inflation  produces a small matter density whose perturbations are not at all adiabatic, the departure from adiabaticity will decay rather than grow as inflation proceeds,
and the departures of the perturbations from their adiabatic values will become exponentially small when the matter density becomes large during reheating.

I was greatly helped in my thinking about this question by a correspondence some months ago with Alan Guth, and more recently by discussions with Christian Armend\'{a}riz-Pic\'{o}n and Eiichiro Komatsu.  This research was supported by the 
National Science Foundation under Grant No. 0071512 and by the Robert A. Welch Foundation and 
also the US Navy Office of Naval Research, Grant No. N00014-03-1-0639, Quantum Optics Initiative. 

\begin{center}
{\bf REFERENCES}
\end{center}

\nopagebreak

\begin{enumerate}

\item A. Guth and S.-Y Pi, Phys. Rev. Lett. {\bf 49}, 1110 (1982); S. Hawking, Phys. Lett.
{\bf 115B}, 295 (1982); A. Starobinsky, Phys. Lett. {\bf 117B}, 175 (1982); F. Bardeen, P. Steinhardt, and M. Turner, Phys. Rev. {\bf D28}, 679 (1983).  A model in which cosmological perturbations arise during inflation from quantum fluctuations in a scalar field derived from the spacetime curvature had been considered earlier by Y. Mukhanov and G. Chibisov, JETP Lett. {\bf 33}, 532 (1981).

\item  H. V. Peiris {\em et al.}, Astrophys. J. Suppl. {\bf 148}, 213 (2003)); P. Crotty, J. Garcia-Bellido, J. Lesgourgues, and A. Riazuelo, Phys. Rev. Lett.
{\bf 91}, 171301 (2003).

\item C. Armend\'{a}riz-Pic\'{o}n, astro-ph/0312389.

\item S. Weinberg, Phys. Rev. {\bf D67}, 123504 (2003).  A somewhat improved discussion is given in the Appendix to S. Weinberg, Phys. Rev. {\bf D 69}, 023503 (2004).  Also see S. Bashinsky and U. Seljak, astro-ph/0310198, especially Appendix B.

\item J. M. Bardeen, P. J. Steinhardt, and M. S. Turner,
Phys. Rev. {\bf D28}, 679 (1983).  This quantity was 
re-introduced by D. 
Wands, 
K. A. Malik, D. H. Lyth, and A. R. Liddle, Phys. Rev. {\bf D62}, 
043527 
(2000).  Also see D. H. Lyth, Phys. Rev. D {\bf 31}, 1792 (1985); K. A. Malik, D. 
Wands, and C. Ungarelli, Phys. Rev. D {\bf 67}, 063516 (2003).

\item J. M. Bardeen, Phys. Rev. {\bf D22}, 1882 (1980);  
D. H. Lyth, Phys. Rev. {\bf D31}, 1792 (1985).  For reviews, see 
J. Bardeen, in {\em Cosmology and Particle Physics}, eds. Li-zhi Fang 
and A. Zee (Gordon \& Breach, New York, 1988); A. R. Liddle and D. H. 
Lyth, {\em Cosmological Inflation and Large Scale Structure} 
(Cambridge 
University Press, Cambridge, UK, 2000).

\end{enumerate}
\end{document}